\documentstyle[psfig,nicV]{article}
\begin{document}
%
\def\msun{M_\odot}
\def\lsun{L_\odot}
%
%
\heading{%
Nucleosynthesis in Intermediate-Mass Asymptotic \\
Giant Branch Stars\\
%
}
\par\medskip\noindent
%
\author{%
John C Lattanzio$^{1,2}$
}
\address{
Department of Mathematics and Statistics, Monash University, Clayton, 3168, 
Victoria, Australia
}
\address{
Institute of Astronomy, Madingley Rd, Cambridge, CB3 0HA, England
}
%
\begin{abstract}
We summarise the main properties of Asymptotic Giant Branch stars,
including their structure, evolution and nucleosynthesis. 
The main physical mechanisms are outlined, as are the uncertainties.
In keeping with the multi-disciplinary nature of this meeting, this
paper is designed for those who are {\bf not experts} in stellar structure.
\end{abstract}
\section{What Is An AGB Star?}
	Most stars will pass through the Asymptotic Giant Branch (AGB) phase.
Although only a brief phase compared to the overall stellar lifetime, 
these stars experience substantial nucleosynthesis during 
this phase. Further, they find ways of mixing the results of this synthesis to
the surface, where stellar winds expel the material into the interstellar
medium. 

	The AGB phase is the last phase of evolution for low and intermediate
mass stars (about $1$--$8\msun$). The evolution leading up to this phase
has been discussed in detail in \cite{JCL_StLouis} and \cite{JCL_Oak}.
Briefly, stars begin by burning hydrogen in their cores, on the main sequence.
Following the exhaustion of their central hydrogen supply, they
become red-giants. During this phase the stars develop a 
deep convective envelope which
extends inwards and can reach into regions where partial hydrogen
burning has taken place. This is called the ``first dredge-up'' event,
and it results in polluting the stellar surface with the products 
of hydrogen burning.

	The ascent of the giant branch is terminated by ignition of the
central helium supply, either degenerately in a core flash (in the case of 
low mass
stars, below about $2.5\msun$) or non-degenerately (for higher masses).
The subsequent evolution is characterized by helium burning in a convective
core and a steadily advancing hydrogen burning shell. For low mass stars
the evolution is complicated by semiconvection, but this a detail we do
not need for our present purposes. The fusion of helium produces $^{12}$C,
and this carbon is in turn subject to alpha capture to form $^{16}$O.
Eventually the helium supply is totally consumed, leaving a core of
carbon and oxygen (the exact proportions of which depend on the infamously
uncertain rate for the $^{12}$C$(\alpha,\gamma)^{16}$O reaction. In any
event, following the exhaustion of central helium the star begins to ascend 
the giant branch again. It is now called the ``second'' or ``asymptotic'' 
giant branch, because of the way the evolutionary tracks seem to approach the
(first) giant branch asymptotically (for some masses, at least). 

	So when a star reaches the AGB it has the structure shown in Figure~1.
If its mass is above about $4\msun$ the shifting of helium burning from
the core to a shell causes the extinguishing of the hydrogen shell.
In lower mass stars this nuclear burning stops the inward advance
of the convective envelope, but in more massive stars the envelope
penetrates the erstwhile hydrogen shell, and the products of hydrogen
burning are again mixed to the stellar surface, in what is known as the
``second dredge-up'' episode. Note that low mass stars do not experience the
second dredge-up.

\begin{figure}
\centerline
{\vbox{\psfig{figure=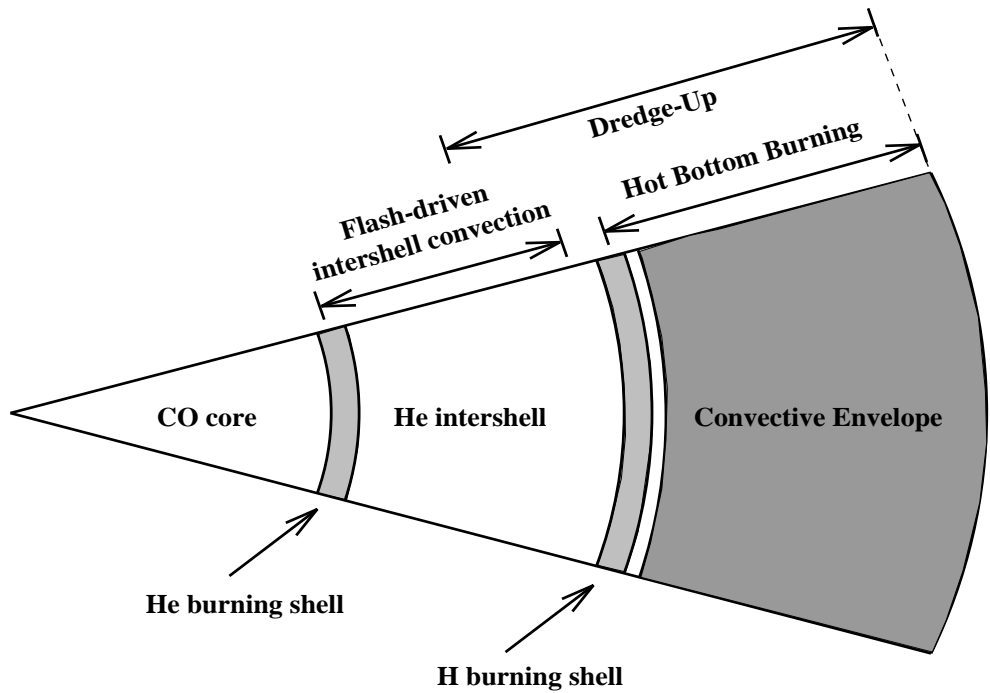,height=9cm,width=\textwidth}}}
\caption[]{\small
Schematic structure of an AGB star.
}
\end{figure}

\section{Why Care About AGB Stars?}
	The importance of AGB stars lies in the nucleosynthesis which occurs
during their evolution along the AGB. During this phase they show an
electron degenerate carbon-oxygen core, and two nuclear shells: one
burning helium and one burning hydrogen. The subsequent nucleosynthesis
comes from two sources:~the thermal instability of the helium shell (known
as ``thermal pulses'' or ``shell flashes'')
and nuclear burning at the bottom of the convective envelope (known as
``hot bottom burning'').

	For most of the evolution on the AGB, the helium shell is largely
inactive. But periodically the shell suffers a thermal runaway and
generates enormous quantities of energy, at rates up to $10^8\lsun$ or so,
for short periods of time (of order a year). This energy release results
in a ``flash-driven convection zone'' which extends from the helium shell
almost to the hydrogen shell, as shown in Figure~1. This mixes the
$^{12}$C produced by the helium shell throughout this convective region.
As the helium flash dies down, the energy deposited in the star causes a
substantial expansion and cooling. The hydrogen shell is extinguished
and the convective envelope reaches in beyond its old location, and
(for later pulses) into the carbon-rich region left behind by the
flash convection. This mixes carbon to the surface of the star, and
is known as the ``third dredge-up''. As the
star now contracts again, the hydrogen shell is re-ignited and the
star begins an extended ``interpulse'' phase (of order $10^4$ years)
during which the hydrogen shell provides essentially all of the
star's luminosity. Eventually there is another flash, and the cycle repeats.

	For stars above about $4\msun$ a second important phenomenon is
the occurrence of hot bottom burning. Here the convective envelope
is so deep that it penetrates into the top of the hydrogen burning
shell, so that nucleosynthesis actually occurs in the envelope of
the star. Temperatures can reach as high as almost $10^8$K. We will
deal with the consequences of these phenomena below.

\section{Nucleosynthesis on the AGB}

\subsection{Thermal Pulses}
	Thermal pulses, and their associated dredge-up, were the
first known site for nucleosynthesis in AGB stars (see, for example,
\cite{Iben75}, \cite{Iben77}). In describing
how thermal pulses work (above) we saw how (primary) $^{12}$C is
produced and mixed to the surface. This is responsible for changing
the star from a normal, oxygen-rich composition, to one which
is carbon-rich (i.e. the number ratio C/O$>1$) and is known as a
Carbon star (spectral type~N; the R and J stars
probably form by a different mechanism). But carbon production is not the
end of the story.

	Arguably the most important synthesis to occur in AGB stars
is the production of $s$-process elements (see \cite{Gallino} for a
more complete summary). This is outlined in Figure~2. Basically, we believe
that somehow there is contact made between the hydrogen-rich envelope
and the carbon-rich intershell zone. The mixing in this region is
partial, leaving behind a varying composition profile. Then, as the star
contracts, proton captures on $^{12}$C produce $^{13}$C. During the
interpulse phase this reaction is followed by $^{13}$C$(\alpha,$n$)^{16}$O
which releases neutrons for capture on iron and similar elements. 

	There are two important aspects of this to discuss. Firstly, until
1995 it was believed that the pocket of $^{13}$C remained in place until
it was engulfed by the next flash-driven convective zone. But in 1995
it was shown by \cite{Oscar} that the $^{13}$C burns {\it in situ\/}
during the interpulse phase. Hence the neutrons are both released and captured
locally, at lower temperatures than found in the convective shell at the
next pulse, and the resulting neutron densities are lower. Studies of the
subsequent $s$-processing assume hydrogen penetrates only a few $10^{-4}\msun$
which is enough to provide a good match to the observations in AGB 
stars (\cite{Gallino98}).

	The second important point is that we are not 
sure exactly how this $^{13}$C
is produced. The problem is that we must mix small amounts of hydrogen
into the carbon-rich region. If there is too much hydrogen then the
CN cycle makes the $^{13}$C that we want, but there are plenty of protons
left for the $^{13}$C to capture another proton to produce $^{14}$N. i.e. we
complete the CN cycle. So it is crucial that there be only small amounts 
of hydrogen present. Initially it was believed that the high opacity 
caused by the recombination of C$^{6+}$ to C$^{5+}$ caused a semiconvective
zone to form, which resulted in the partial mixing of the two 
zones (\cite{IR1}). Another mechanism has been outlined by \cite{Falk}, who
has shown that a combination
of convective overshooting and partial mixing (implemented via a diffusion 
equation) can produce both extensive dredge-up and 
a $^{13}$C pocket. This is very promising, although it does introduce 
another parameter which must be calibrated somehow. Another possibility 
is that shear mixing could occur at the bottom of the convective envelope.
For rotating stars we would expect a significant shear layer at the
bottom of the convective envelope, which is exactly where this mixing must 
occur. In summary, it seems 
clear that some hydrogen is mixed into the intershell region, but we are
still unsure how.

\begin{figure}
\centerline
{\vbox{\psfig{figure=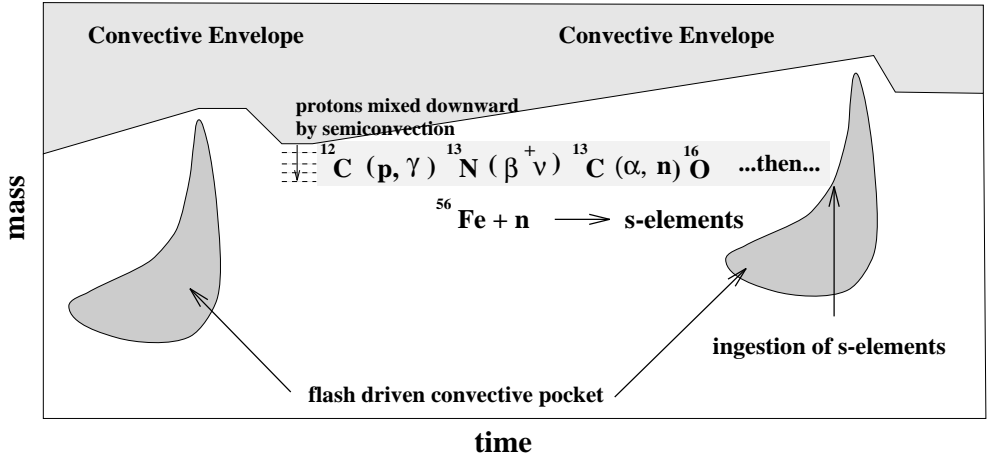,width=\textwidth}}}
\caption[]{\small
Schematic showing how $s$-process elements are produced in AGB stars.
}
\end{figure}

AGB stars can also make $^{19}$F through thermal pulses, although the
details are rather complicated. There is a delicate interplay between
many reactions, primarily concerning the fate of the abundant $^{14}$N and
the production of $^{15}$N in
the flash-driven convection zone (\cite{Fetal},\cite{Mowlavi1}). 
The basic production scheme is shown in
Figure~3.  Some $^{13}$C~produces neutrons via the $^{13}$C$(\alpha,
$n$)^{16}$O~reaction described above, and some of these neutrons
are captured by $^{14}$N~to produce $^{14}$C and protons. These protons,
plus possibly some from $^{26}$Al$($n,p$)^{26}$Mg, are then captured by
$^{18}$O~and the sequence
$^{18}$O(p$,\alpha)^{15}$N$(\alpha,\gamma)^{19}$F~produces 
the observed $^{19}$F, which is then dredged to the surface  in the
usual way following the next pulse. Note that the neutron captures on
$^{14}$N must compete with $\alpha$-captures, and likewise the production of
$^{15}$N is 
very dependent on the temperature (and hence the stellar mass).
It appears that the amount of
$^{13}$C left over from hydrogen burning is dramatically short of
the amount required for $^{19}$F production (\cite{Mowlavi2}) and hence 
this is further evidence for the existence of a $^{13}$C pocket.

\begin{figure}
\centerline
{\vbox{\psfig{figure=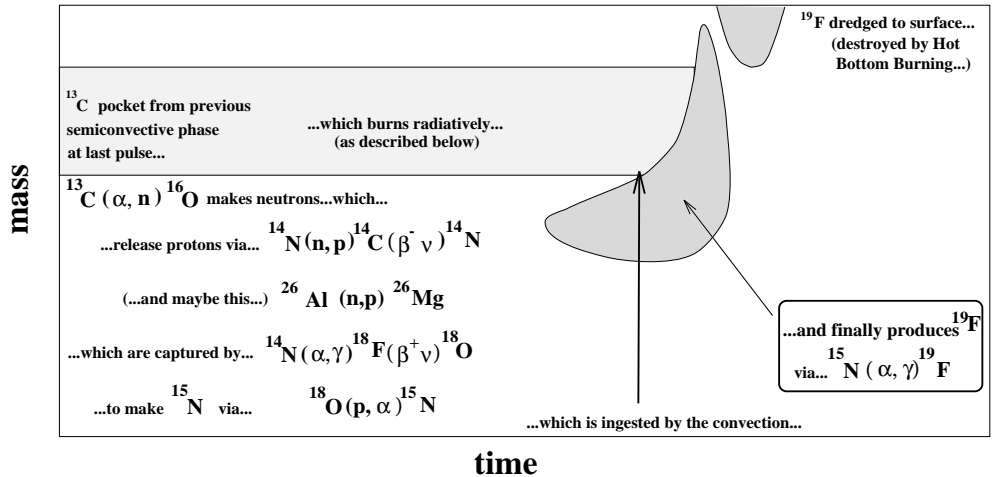,width=\textwidth}}}
\caption[]{\small
Schematic of the production of $^{19}$F during thermal pulses on the AGB.
}
\end{figure}

\subsection{Hot Bottom Burning}
	For stars more massive than about $4\msun$ the temperature at the
bottom of the convective envelope becomes high enough for some
reactions to occur. The first effect of this ``hot bottom burning''
(hereafter HBB) is
that the star produces $^7$Li via the Cameron-Fowler mechanism.
At the bottom of the envelope, $^3$He
(left over from earlier hydrogen burning on the main-sequence) captures
an alpha particle to make $^7$Be, which decays into $^7$Li. However,
at the high temperatures necessary for the first step of this reaction, 
the $^7$Be is destroyed by the PPIII chain, and the $^7$Li itself is
destroyed by the PPII chain. So if the abundance of $^7$Li is
to increase, we must produce the $^7$Be and then mix it away to regions
of lower temperature, where it can decay into $^7$Li. Detailed
calculations by \cite{BS92}
showed that $^7$Li
was produced as soon as the temperature at the bottom of the
convective envelope exceeded $50\times 10^6$K, and produced abundances
up to $\log\varepsilon(^7$Li$) \sim 4.5$ in stars with
$M_{bol} \simeq -6$ to~$-7$ (note that
$\log\varepsilon(^7$Li$) \equiv \log\{n(^7$Li$)/n($H$)\}+12$).
This is in excellent agreement with the observations
(\cite{SL89} and  \cite{SL90}).

	The next consequence of HBB is that the $^{12}$C which has been
added to the envelope after each dredge-up event is now processed by
the CN cycle into $^{13}$C and $^{14}$N. It turns out that the entire envelope
is processed many times during the interpulse phase, and the surface
ratio of $^{12}$C/$^{13}$C can reach its equilibrium value of about 3-3.5.
Further, as $^{12}$C is destroyed in the process, the star can be prevented
from becoming a C-star by HBB (\cite{BSA}), and it can even produce significant
amounts of (primary) $^{14}$N by this process (\cite{bright}).

	Two other important consequences of HBB are the production of
$^{26}$Al and the destruction of $^{19}$F (\cite{JCL_StLouis}). 
This is important
for our understanding of meteorite grains (see \cite{grains}) and for
the overall galactic enrichment of $^{19}$F, but we will defer the 
details to elsewhere.

\subsection{But$\ldots$}
	But this picture
has to be modified slightly. Both HBB and dredge-up require a reasonably
large envelope mass. Until recently we did not know how large.
The continued effect of mass-loss on the AGB is to reduce the envelope mass.
So both dredge-up and HBB will eventually cease, before the stars leaves the
AGB as a planetary nebula. But which ceases first? Dredge-up or HBB?
It was recently shown by \cite{bright} that
HBB is the first to end, while dredge-up continues. Hence
the carbon continues to be added to the envelope without it being
burned, and the star does indeed become a carbon star near the end of
its life, despite earlier HBB. At this time, however, the mass loss rate is so
large that the star is not visible optically, which explains recent
infra-red surveys which show that of order of 50\% of the obscured
AGB stars are in fact C-stars (\cite{loon}).

\section{Do We Really Understand It All?}
	Lest the reader think that all is understood, there are
some significant uncertainties in the details, which I have not discussed
at all. Some of these include the form of the mass-loss, the value
of the mixing length, various reaction rates, as well as the ever-present
(and mostly ignored) unknown effects of rotation. But the one uncertainty
which I do wish to elaborate on (slightly) is that of dredge-up.

	There has been, and continues to be, a large disparity in the
extent (or even occurrence) of dredge-up found by various
authors. The importance of differences in numerical treatment of
convection and mixing was shown by \cite{FL96}, which partly explains
some of the different results which appear in the literature. The
details depend on how one implements the mixing of convective zones within
an iteration scheme, on the one hand, and whether one naively uses
the Schwarzschild criterion for determining the convective
boundary or whether one allows the model to find a buoyantly neutral
edge. Exactly how one performs the mixing can have large
effects too (\cite{PRW}). 

	The recent work by \cite{Falk} on overshooting is promising, but
does suffer from the (necessary) inclusion of another unknown parameter.
Their models show deep dredge-up, which saturates at a level independent
of further changes of the numerical parameters (\cite{FalkGrenoble}) which is
consistent with calculations by \cite{Mowlavihere}. Much work remains
to be done on this complex problem, which is basically a multi-dimensional
time-dependent hydrodynamical one. It is really no wonder 
that our one dimensional
hydrostatic models are in trouble!

\section{Conclusions}
	Table~1 lists the main elements involved in nucleosynthesis in
AGB stars and indicates where they are made or destroyed. This (rough) summary
explains the interest in AGB stars: although a short phase they are
common and of great importance to the chemical enrichment of the galaxy.

	But there are still more things to learn about these stars.
The recent possibility of ``degenerate pulses'' (\cite{dp}) has yet
to be investigated fully. AGB stars exhibit much complex physics,
and we should not be so complacent as to think that we have
uncovered most of their workings yet!

%
%

\begin{center}
\begin{tabular}{c c c}
\multicolumn{3}{l}{{\bf Table 1.} Main nucleosynthesis results on the AGB } \\
\\
\hline\\
Element&Thermal Pulses&Hot Bottom Burning\\%
\\%
\hline
\\
$^{12}$C  & produced & destroyed \\
$^{13}$C  & little effect  & produced \\
$^{14}$N  & little effect  & produced \\
$^{15}$N  & little effect  & destroyed \\
$^{16}$O  & little effect  & destroyed \\
$^{17}$O  & little effect  & produced \\
$^{18}$O  & little effect  & destroyed \\
$^{19}$F  & produced  & destroyed \\
$^{26}$Al & slight production&large production\\
$s$-elements & produced  & little effect \\
\\
\hline
\end{tabular}
\end{center}
%

%
\vfill

\begin{iapbib}{99}{

\bibitem{grains}Bernatowicz, T. J., and Zinner, E., Eds., 1997, {\sl Astrophysical 
Implications of the Laboratory Study of Presolar Materials\/}, (New York: AIP).

\bibitem{BS92}Boothroyd, A. I., and Sackmann, I.-J., 1992,
	\apj, 392, L71

\bibitem{BSA}Boothroyd, A. I., Sackmann, I.-J., and Ahern, S. C., 1993,
	\apj, 416, 762

\bibitem{Fetal}Forestini, M., {\it et al\/}, 1992, \aeta, 261, 157

\bibitem{FL96}Frost, C. A., and Lattanzio, J. C., 1996, \apj, 473, 383

\bibitem{dp}Frost, C. A., Lattanzio, J. C., and Wood, P. R., 1998,
	\apj, 500, 355

\bibitem{bright}Frost, C. A., {\it et al\/}, 1998, \aeta, 332, L17

\bibitem{Gallino98}Gallino, R., {\it et al\/}, 1998, \apj, 497, 388

\bibitem{Gallino}Gallino, R., Busso, M., and Lugaro, M., in {\it
Astrophysical Implications of the Laboratory Study of Presolar Materials\/},
eds Bernatowicz, T. J., and Zinner, E., (New York: AIP) p115

\bibitem{FalkGrenoble}Herwig, F., presentation at {\it $3^{rd}$ Torino Workshop on 
Nuclear Astrophysics\/}, Grenoble, September 1998.

\bibitem{Falk}Herwig, F., Bl\"ocker, T., Sch\"onberner, D and El Eid, M., 1997,
	\aeta, 324, 81

\bibitem{Iben75}Iben, I., Jr., 1975, \apj, 196, 252

\bibitem{Iben77}Iben, I., Jr., 1977, \apj, 217, 788

\bibitem{IR1}Iben, I., Jr., and Renzini, A., 1982, \apj, 263, L23

\bibitem{JCL_Oak} Lattanzio, J. C., 1998, in {\it
Second Oak Ridge Symposium on Atomic and Nuclear Astrophysics\/}, IOP, p299

\bibitem{JCL_StLouis} Lattanzio, J. C., and Boothroyd, A. I., 1997, in {\it
Astrophysical Implications of the Laboratory Study of Presolar Materials\/},
eds Bernatowicz, T. J., and Zinner, E., (New York: AIP), p.85

\bibitem{loon}van Loon, J. Th., {\it et al\/}, 1998, \aeta, 329, 169

\bibitem{Mowlavihere}Mowlavi, N., this volume.

\bibitem{Mowlavi1}Mowlavi, N., Jorissen, A., and Arnould, A., 1996, 
	\aeta, 311, 803

\bibitem{Mowlavi2}Mowlavi, N., Jorissen, A., and Arnould, A., 1998, 
	\aeta, 334, 153

\bibitem{SL89} Smith, V. V., and Lambert, D. L., 1989, \apj, 345, 375

\bibitem{SL90} Smith, V. V., and Lambert, D. L., 1989, \apj, 361, L69

\bibitem{Oscar}Straniero, O., {\it et al\/}, 1995, \apj, 440, L85

\bibitem{PRW}Wood, P. R., 1981, \apj, 248, 311
}
\end{iapbib}
\end{document}